\documentclass[twocolumn,showpacs,amsmath,amssymb,superscriptaddress]{revtex4}
\usepackage{amsmath,amssymb,color,latexsym,natbib,graphicx}
\input{psfig}

\usepackage{dcolumn}
\usepackage{bm}

\def\eg{{\it e.g.}\ } 
 
\def\ie{{\it i.e.}\ }
 
\def\kpe{k_{\perp}} 
\def\bB0{{\bf B}_0}
\def\ep{{\bf e_{\parallel}}}
\def\kpa{k_{\parallel}}
\def\zs{{\bf z_{\perp}}}
\def\zp{{\bf z_{\parallel}}}
\def\zsf{{\bf {\hat z}_{\perp}}}
\def\zpf{{\bf {\hat z}_{\parallel}}}
\def\alb{\bar \alpha}
\def\beb{\bar \beta}

\def\be{\begin{equation}}
\def\ee{\end{equation}}

\begin{document}

\preprint{1}

\title{Energy decay laws in strongly anisotropic MHD turbulence}

\author{Barbara Bigot}
\affiliation{Institut d'Astrophysique Spatiale (IAS), B\^atiment 121, 
F-91405 Orsay (France); \\
Universit\'e Paris-Sud XI and CNRS (UMR 8617)}
\affiliation{Laboratoire Cassiop\'ee, UMR 6202, OCA, BP 42229, 06304 Nice Cedex 4, France}

\author{S\'ebastien Galtier}
\affiliation{Institut d'Astrophysique Spatiale (IAS), B\^atiment 121, F-91405 Orsay (France); \\
Universit\'e Paris-Sud XI and CNRS (UMR 8617)}

\author{H\'el\`ene Politano}
\affiliation{Laboratoire Cassiop\'ee, UMR 6202, OCA, BP 42229, 06304 Nice Cedex 4, France}

\date{\today}

\begin{abstract}
We investigate the influence of a uniform magnetic field $\bB0=B_0 \ep$ on
energy decay laws in incompressible magnetohydrodynamic (MHD) turbulence.
The nonlinear transfer reduction along $\bB0$ is included in a model that
distinguishes parallel and perpendicular directions, following a phenomenology {\it \`a la}
Kraichnan. We predict a slowing down of the energy decay due to anisotropy in the limit
of strong $B_0$, with distinct power laws for energy decay of shear- and
pseudo-Alfv\'en waves. Numerical results from the kinetic equations of Alfv\'en wave 
turbulence recover these predictions, and MHD numerical results clearly tend to follow 
them in the lowest perpendicular planes.
\end{abstract}


\maketitle

Predicting the evolution of freely decaying turbulence remains one of the most difficult 
problems in turbulence. For the sake of simplicity, this problem is often tackled under the assumption 
of homogeneity and isotropy. In the case of 3D Navier-Stokes flows, the Kolmogorov 
prediction for the kinetic energy decay law thus reads $E_v(t) \sim (t-t_*)^{-10/7}$, where 
$t_*$ is the time origin for power law decay \cite{K41}. Close power law indices are measured 
in grid turbulence experiments although boundary effects may alter the result \cite{comte}. 
The generalization of decay laws to other situations is still currently under debate. For example in 
rotating turbulence, a recent experiment shows that neutral fluids behave differently from pure 
(non rotating) flows with a slowing down of the kinetic energy decay due to the presence 
of anisotropy \cite{morize}. In this case, the original isotropic Kolmogorov description has to be 
modified to include rotation effects that may lead to strong spectral anisotropy (see \eg Ref. 
\cite{galtier03}). 

In this Letter, we investigate the influence of an external uniform magnetic field on the 
energy decay laws in freely incompressible magnetohydrodynamic (MHD) turbulence. The MHD  
approximation has proved to be quite successful in the study of a variety of astrophysical plasmas 
like those found in the solar corona, the interplanetary medium or in the interstellar clouds. These 
media are characterized by extremely large Reynolds numbers (up to $10^{13}$) \cite{Tajima} 
with a range of available scales from $10^{18}$m to few meters. The isotropy assumption is 
particularly difficult to justify when dealing with astrophysical flows since a large scale magnetic 
field is almost always present like in the inner interplanetary medium where the magnetic field lines 
form an Archimedean spiral near the equatorial plane (see \eg Ref. \cite{galtier06}). The present 
study, although theoretical, appears therefore particularly important to extract some universal 
features of turbulent plasmas. The MHD equations in presence of an external uniform magnetic 
field $\bB0$$=B_0 \ep$ read  
\be
\partial_t {\bf v} - B_0 \partial_{\parallel} {\bf b} + {\bf v} \cdot \nabla \, {\bf v} = 
- {\bf \nabla} P_* + {\bf b} \cdot \nabla \, {\bf b} + \nu \Delta {\bf v} \, ,
\label{mhd1}
\ee
\be
\partial_t {\bf b} - B_0 \partial_{\parallel} {\bf v} + {\bf v} \cdot \nabla \, {\bf b} = 
{\bf b} \cdot \nabla \, {\bf v} + \eta \Delta {\bf b} \, , 
\label{mhd3}
\ee
with $\nabla \cdot {\bf v} = 0$ and $\nabla \cdot {\bf b} = 0$. 
The magnetic field ${\bf b}$ is normalized to a velocity (${\bf b} \to \sqrt{\mu_0 n m_i} 
\, {\bf b}$, with $m_i$ the ion mass and $n$ the electron density), ${\bf v}$ is the plasma flow 
velocity, $P_*$ the total (magnetic plus kinetic) pressure, $\nu$ the viscosity and $\eta$  
the magnetic diffusivity. The role of the $\bB0$-field on the flow behavior has been widely 
discussed in the community (see \eg Ref. \cite{strauss76,shebalin,
goldreich95,ng96,matthaeus98,galt00,milano2001,chandran}). 
One of the most clearly established results is the bi-dimensionalization of 
an initial isotropic energy spectrum with a strong reduction of nonlinear transfers along $\bB0$. 

In the past, several papers have been devoted to predictions of energy decay laws in isotropic 
MHD turbulence \cite{hatori,politano89,biskamp,kinney,galtier97}. 
We first review one of them \cite{galtier97} which follows the Kolmogorov-Kraichnan spirit that 
we then adapt to strongly anisotropic flows. In the simplest case (balance turbulence), the 
derivation of self-similar decay laws in MHD relies on few basic assumptions which are (i) a 
weak correlation between velocity and magnetic fields 
(which allows to use below the variable $z$ instead of $z^{\pm}$, where 
${\bf z}^{\pm}={\bf v} \pm {\bf b}$), 
(ii) a power law spectrum in $E(k) \sim k^s$ (with $s=D+1$, $D$ being the space dimension) 
for the low wavenumber energy, \ie at scales 
larger than the integral scale $\ell$ from which the inertial range begins, and, obviously (iii) a power 
law time dependence for $E(t) \sim (t-t_*)^{-\alpha}$ and $\ell(t) \sim (t-t_*)^{\beta}$, where 
$\alpha$ and $\beta$ are two unknown indices. Another hypothesis, directly related to the 
invariance of the Loitsianskii integral \cite{davidson}, tells that the modal spectrum scales like 
$k^{s-2}$ at low wavenumbers and, with sufficient scale separation, this dominates the total energy so that 
$E \sim \ell^{-(s+1)}$, hence the first relation $\alpha = \beta(s+1)$. A second relation may be obtained 
from the energy transfer equation $\epsilon = -dE/dt \sim E/\tau_{tr} \sim (t-t_*)^{-\alpha-1}$, where 
$\tau_{tr}$ is the transfer time and $\epsilon$ is the transfer rate. According to the Iroshnikov-Kraichnan 
(IK) phenomenology \cite{iro}, $\tau_{tr} = \tau_{NL}^2/\tau_{A}$, with 
$\tau_{NL}=\ell / z_{\ell}$, the eddy turnover time, and $\tau_{A} = \ell / B_0$, the Alfv\'en time. 
Substituting previous times into the energy transfer equation leads to $1=\alpha+\beta$. Finally one 
obtains $\alpha = (s+1)/(s+2)$ and $\beta=1/(s+2)$. The predictions for 3D MHD turbulence ($s=4$) 
are then $E(t) \sim (t-t_*)^{-5/6}$ and $\ell(t) \sim (t-t_*)^{1/6}$. The energy decay law for isotropic 
MHD turbulence has been favorably compared with direct numerical simulations, in particular in the 
2D case for which higher Reynolds numbers can be reached than in the 3D case \cite{kinney,galtier97}. 
Note that other phenomenologies exist for explaining, in particular, the influence of magnetic helicity 
on the decay laws \cite{Biskamp1999}, a situation different from here where it is negligible. 

We now derive an anisotropic version of the previous heuristic description which is well 
adapted to MHD flows permeated by a strong uniform magnetic field. The main characteristic of 
such flows is that nonlinear transfers are strongly damped along the mean field $\bB0$ direction, 
leading preferentially to perpendicular ($\perp$) transfers. In Fourier space, this means that the modal 
energy spectrum mainly develops a power law scaling in $\kpe$-wavenumbers whereas the 
scaling along parallel wave numbers, $\kpa$, does not change very much. In practice, we assume 
that the intensity $B_0$ is strong enough to ignore parallel transfers and we make the approximation
$k \sim \kpe \gg  \kpa$ which means, in particular, that the initial condition of the system is confined 
to the largest parallel scales. In this situation, it is straightforward to show that shear-Alfv\'en wave
energy corresponds to perpendicular fields ($\zs$) whereas pseudo-Alfv\'en wave energy mainly comes 
from parallel fields ($\zp$). Our heuristic description is focused on shear-Alfv\'en waves, 
and not pseudo-Alfv\'en waves, since it is well known that under such anisotropic configuration, the 
latter are slaved to the former \cite{goldreich95,galt00}. 

We make three equivalent hypotheses as in the isotropic model and we note $E_{\perp}$, 
$\ell_{\perp}$ and $\ell_{\parallel}$ the shear-Alfv\'en wave energy, the perpendicular and parallel integral 
scales respectively. The $\ell_{\parallel}$ length scale, although appearing sometimes in our derivation, is 
assumed time independent because of the strong nonlinear anisotropy. In that case,  we assume (a) 
$v_{\perp} \sim b_{\perp} \sim z_{\perp}$, (b) for $\kpa$ fixed, $E_{\perp} (\kpe,\kpa) \sim \kpe^{D}$ 
(where $D$ is the space dimension)
at low perpendicular wavenumbers, \ie at scale larger than $\ell_{\perp}$, and (c) 
$E_{\perp}(t) \sim (t-t_*)^{-\alb}$ and $\ell_{\perp} \sim (t-t_*)^{\beb}$, where $\alb$ and $\beb$ are the 
new unknown indices. The second assumption means a $D-1$ power law index for the modal spectrum 
before the integral $\kpe$-wavenumber; furthermore it leads to estimate 
$E_{\perp}(t) \sim \ell_{\perp}^{-(D+1)} \ell_{\parallel}^{-1}$, and hence the (same) first relation 
\be
\alb = \beb(D+1) \, .
\label{equ1}
\ee
A second relation may be obtained by using the energy transfer equation 
\be
\epsilon_{\perp} = -dE_{\perp}/dt \sim E_{\perp}/\tau_{tr} \sim (t-t_*)^{-\alb-1} \, ,
\label{equ2}
\ee
where $\tau_{tr} = \tau_{NL}^2/\tau_{A}$, with, for anisotropic transfers, 
$\tau_{NL}= 1 / (\kpe {z_{\perp}}_{\ell_{\perp}})$ and $\tau_{A} = 1 / (\kpa B_0)$.
Substituting these times into (\ref{equ2}) leads to the (new) relation 
\be
1=\alb+2\beb \ .
\label{equ3}
\ee
Finally, (\ref{equ1}) and (\ref{equ3}) lead to two new scaling exponents:
\be
\alb = {D+1 \over D+3} \, ; \, \,  \beb={1 \over D+3} \, .
\label{equ4}
\ee
Hence the predictions for 3D anisotropic MHD turbulence ($D=3$) $E_{\perp}(t) \sim (t-t_*)^{-2/3}$ and 
$\ell_{\perp}(t) \sim (t-t_*)^{1/6}$. These results show, in particular, a slowing down of the energy decay for 
shear-Alfv\'en waves compared to the total energy in the isotropic case (where no distinction is made
between shear- and pseudo-Alfv\'en waves). 

The extension of our approach to pseudo-Alfv\'en waves is not direct since these waves are (mainly) 
slaved to shear-Alfv\'en waves. 
We remind that shear-Alfv\'en and pseudo-Alfv\'en waves are the two kinds of linear perturbations 
about the equilibrium, the latter being the incompressible limit of slow magnetosonic waves. 
For the pseudo-waves, the heuristic description based on nonlinear 
transfers is misleading. Instead, it seems suitable to find a relationship between the different type of 
waves. The divergence free condition provides this relation which eventually leads to a prediction 
for the energy decay law. In Fourier space, the divergence free condition reads 
\be
{\bf \kpe} \cdot \zsf + {\bf \kpa} \cdot \zpf = 0 \, 
\label{equ6}
\ee
where $\zsf$ and $\zpf$ are the Fourier transform of the cartesian fields which may be associated, 
under strong anisotropy assumption, mainly to the shear- and pseudo-Alfv\'en waves respectively. Since 
we assume a weak cross-correlation (balance turbulence), it is not necessary to introduce the Els\"asser 
variables. Simple manipulations lead to $\kpe^2 E_{\perp} \sim \kpa^2 E_{\parallel}$, 
where $E_{\parallel}$ denotes the pseudo-Alfv\'en wave energy. Since nonlinear transfers along 
the $\bB0$-direction are negligible, $\kpa$ may be seen as a mute variable. Therefore, the energy 
decay law for pseudo-Alfv\'en waves should be 
\be
E_{\parallel} \sim (t-t_*)^{-1} \, . 
\label{equ8}
\ee
Our anisotropic model thus predicts quite different energy decay laws for shear- and pseudo-Alfv\'en
waves, the latter not depending on the system dimensions. 
An important issue concerns the balance $\kpa \sim \kpe^{2/3}$ law found in many simulations 
(see \eg \cite{cho}). 
Our analysis is based on the permanence of big eddies \cite{K41} which is directly 
linked to the conservation of the spectral scaling law at the largest scales, \ie at scales larger than the 
ones of the inertial range where the "critical balance" is well observed. The 2/3 law is {\it a priori} not 
included in the model (as well as the exact scaling law for the energy spectrum) because the assumption 
of the existence of an inertial range where energy is evacuated from the reservoir is enough. 
Therefore, the integral length scales $\ell_{\perp}$ and $\ell_{\parallel}$ are not linked by the 2/3 law even 
in the derivation of equation (\ref{equ3}) since we are only dealing with scales before the inertial range.
The possible persistence of the 2/3 law at the largest scales is nevertheless an important issue 
which, in principle, may be investigated from the derivation of the von K\'arm\'an-Howarth equation 
in anisotropic MHD turbulence, from which a Loitsianskii type invariant could be found.
This is a huge challenge for the future which is, of course, out of the scoop of this paper.

\begin{figure}[h]
\centerline{\hbox{\psfig{figure=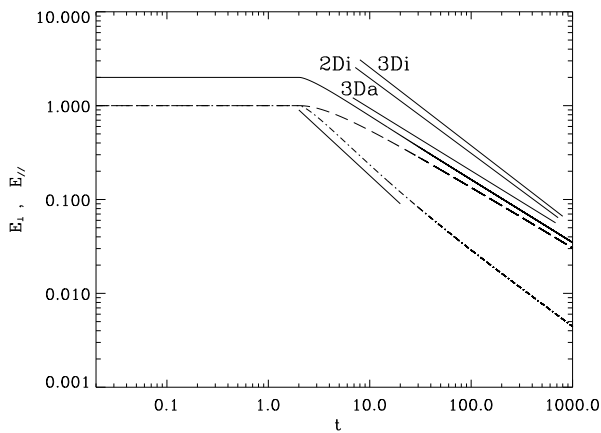,width=8.cm,height=6.2cm}}}
\caption{Temporal decays of shear-Alfv\'en (long-dash), pseudo-Alfv\'en 
(dash-dot) and total (solid) energies, together with three theoretical slope predictions: 3D isotropic (3Di) in 
$t^{-5/6}$, 2D isotropic (2Di) in $t^{-4/5}$ and 3D anisotropic (3Da) in $t^{-2/3}$.  
A $t^{-1}$ slope is also ploted for comparison with the pseudo-Alfv\'en energy decay.}
\label{Fig1}
\end{figure}

In order to check the model validity, we perform numerical simulations of strongly anisotropic MHD flows. 
The first data set comes from integrations of the kinetic equations of Alfv\'en wave turbulence 
derived and analyzed in \cite{galt00}. This regime describes the asymptotic limit of strong $B_0$. 
We do not re-derive or even re-write these equations that the reader can find in \cite{galt00} 
(equations (54) and (55)). One of the main results found in this regime is a $\kpe^{-2}$ scaling 
law for the energy spectrum (for both shear- and pseudo-Alfv\'en waves) which is an 
exact power law solution of the wave kinetic equations (in absence of cross-correlation). Another 
important result is the total absence of parallel transfer. For that reason, the numerical simulations are 
made at a fixed $\kpa$. A non-uniform grid in Fourier space (see for details \cite{galt00}) is used to 
achieve a highly turbulent state at a Reynolds number of about $10^5$, with $\nu=\eta=2 \times 10^{-5}$.
Figure \ref{Fig1} displays the time evolution of the shear-Alfv\'en, pseudo-Alfv\'en and total energies 
in absence of cross-correlation. 
\begin{figure}[h]
\centerline{\hbox{\psfig{figure=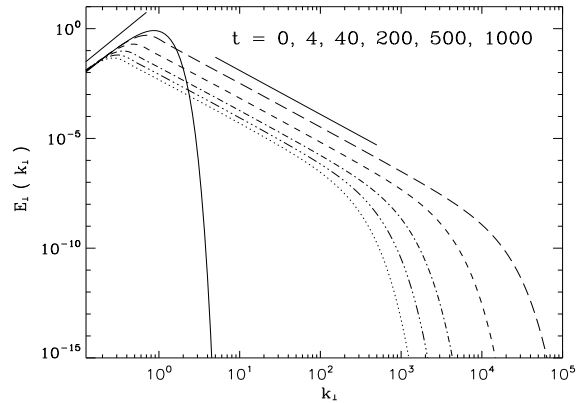,width=8.cm,height=5.8cm}}}
\caption{Temporal evolution of shear-Alfv\'en energy spectra. A $\kpe^3$ and 
$\kpe^{-2}$ power laws are ploted for comparison.}
\label{Fig2}
\end{figure}
We first note the existence of a transient period during which the energy is conserved. During this 
period the energy cascades towards smaller scales following a $\kpe^{-7/3}$ scaling law rather than 
the $\kpe^{-2}$ exact solution \cite{galt00}.
A clear power law behavior appears for both types of waves. These power laws are in agreement with 
the theoretical predictions made for 3D anisotropic flows (for comparison, the predicted slopes 
for the 3D and 2D isotropic cases are also given). Surprisingly, the total energy follows very precisely 
the theoretical prediction over more than two decades whereas pseudo-Alfv\'en waves decay slower 
than $t^{-1}$ at very large times. This discrepancy may be linked to the saturation of the integral
length scale as it can be observed from  the temporal evolution of the shear-Alfv\'en wave spectra 
(Fig. \ref{Fig2}). Indeed, a self-similar energy decay is observed with a slow increase of the integral 
length scale that can be roughly estimated 
from the wavenumber at which the inertial range begins (\ie the maxima of the spectra). At later times, 
this scale is close to the maximum size of the numerical box whereas the large-scale power law in $\kpe^3$ 
is still preserved. The large scales are even more reduced for (slaved) pseudo-Alfv\'en waves since 
they decay faster. Actually, the finite size box effect may explain the change of decay law at very large 
times for pseudo-Alfv\'en wave energy. 
\begin{figure}[h]
\centerline{\hbox{\psfig{figure=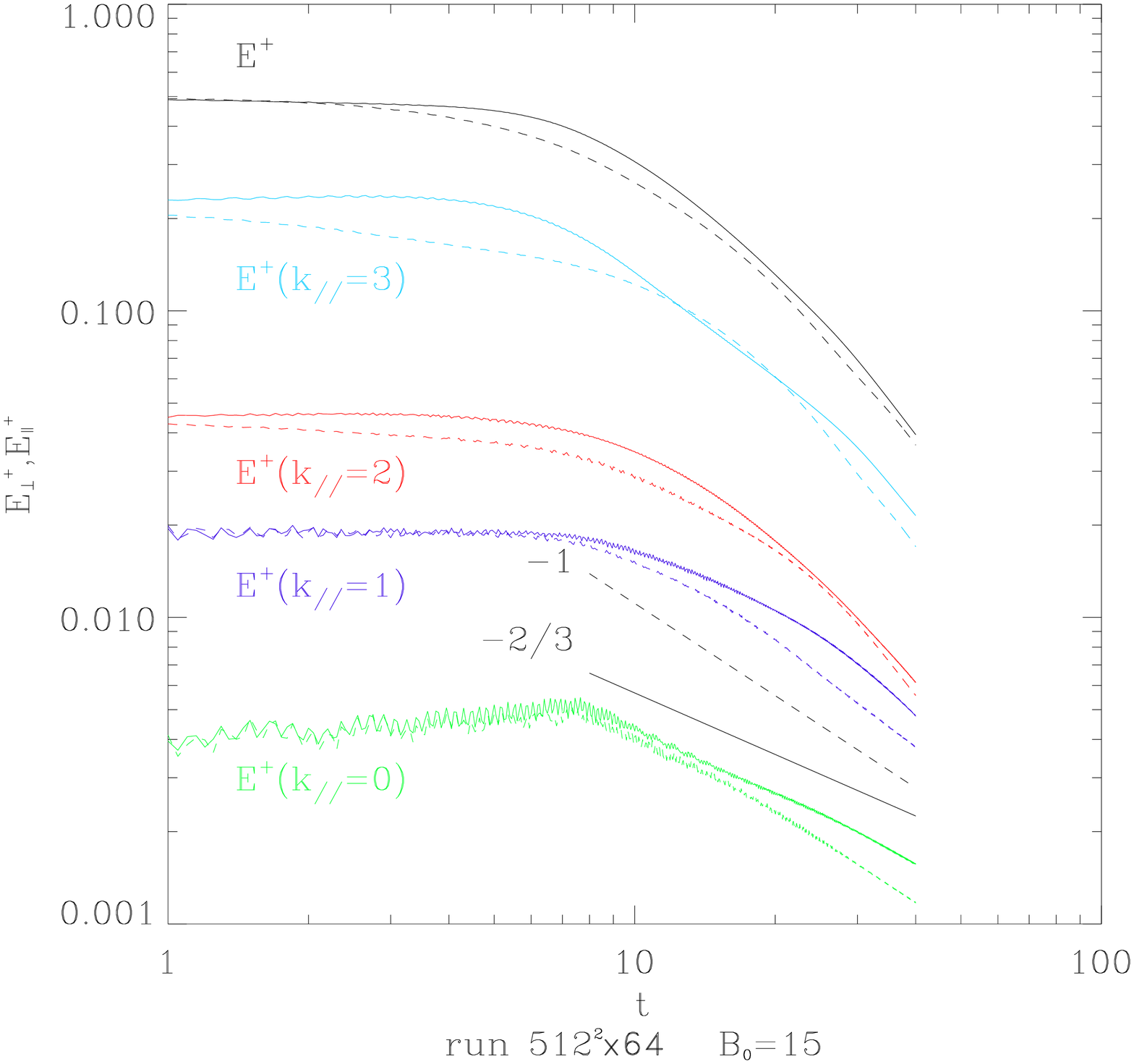,width=8.cm,height= 6.2cm}}}
\caption{Evolution of shear- (solid lines) and pseudo-Alfv\'en (dash lines) wave spectral energies 
at fixed $\kpa$ wavenumber, together with the total energies. Note that curves at 
$\kpa=0$ and $\kpa=1$ are shifted by a factor of 2 for clarity. (Color online).}
\label{Fig3}
\end{figure}
Direct numerical simulations of 3D incompressible MHD equations (\ref{mhd1}--\ref{mhd3}) are also 
performed with a pseudo-spectral code including de-aliasing. 
A high resolution with $512^2$ in the $\bB0$-transverse planes whereas only $64$ grid points are 
taken in the longitudinal direction. Such a situation was analyzed to explore the self-consistency of the 
reduced MHD model \cite{Oughton04} with the conclusion that small values of viscosities, adjusted 
according to the transverse dynamics, are not incompatible with the smaller spatial resolution in 
the longitudinal direction since the transfer towards small scales is also reduced along the uniform 
magnetic field. We have checked that viscosity values,  $\nu=\eta=5 \times 10^{-4}$, are indeed well 
adjusted \cite{rot}. The initial condition corresponds to a modal energy spectrum in agreement with 
the phenomenology described above, with a $D=3$ power law at largest scales : 
$E^\pm(\kpe,\kpa) = C(\kpa) \kpe^3$ for $\kpe$ and $\kpa$ $\in [0,4]$, the value of $C(\kpa)$ increasing 
with $\kpa$ to reach a maximum at $\kpa=4$. This initial spectrum 
allows a transient period of cascade towards smaller scales during which energy is mainly 
conserved.  The uniform magnetic field is fixed to $B_0=15$. 
Initially, the ratio between kinetic and magnetic energies is one, whereas the correlation
$2\langle {\bf v \cdot b} \rangle/\langle \vert {\bf v}\vert^2 + \vert{\bf b}^2 \vert \rangle$
is zero (remaining less than $4\%$ up to $t=40$).
In Fig. \ref{Fig3}, time evolutions of shear- and pseudo-Alfv\'en wave energies are ploted at fixed parallel 
wavenumbers, namely $\kpa=0$, $1$,  $2$, $3$, together with total energies 
(as obtained by integrating  $E_{\parallel}$ over all $\kpa$).
Note that we chose to show the behavior of one type of waves, \ie $E_{\perp, \parallel}^+$ 
 (with by definition $E^{\pm} = \langle \vert {\bf z}^{\pm} \vert^2 \rangle/2$), 
as $E_{\perp, \parallel}^-$ energies behave the same way. 
We observe that systematically $E_{\perp}^+ > E_{\parallel}^+$ at the final 
time which confirms the previous results. A clear tendency towards power law behavior is also found, 
in particular at low $\kpa$, with exponents close to the heuristic predictions. This was already observed 
at smaller resolution (not shown) and seems to be a general tendency of anisotropic MHD flows. 
However, this is less clear in planes at $\kpa=2$ and $3$, probably due to enhanced dissipations there 
(for example, at $t=20$, $\sim 60\%$ of the shear wave energy is lost at $\kpa=2$, and $\sim 73\%$ at 
$\kpa=3$) leading thus to shortened self-similar decay ranges. 
Actually, this remark may explain the power law steepening of the energy decay, when integrated over 
all $\kpa$, since the energy loss is even more pronounced in
higher $\kpa$-planes. This average effect has never been emphasized in the literature
but seems to be the most important obstacle to see the decay laws as well as the spectral laws
predicted by anisotropic theories like wave turbulence \cite{Bigot07}.
Note that other initial conditions with an extended energy spectrum in $\kpa$ might reduce the 
discrepancy found between the model and the simulation at high $\kpa$, a situation not investigated 
here because of the constraint due to the reduced parallel resolution.

In this Letter, we show the influence of a uniform magnetic field on energetic decays in MHD 
turbulence. Modified self-similar laws are derived, with a $t^{-2/3}$ and a $t^{-1}$ decay, respectively 
for shear- and pseudo-Alfv\'en waves, under strong anisotropy. To our knowledge, this decay 
analysis is the first in the context of wave turbulence and could be extended to many other problems.
Integrations of kinetic equations of Alfv\'en wave turbulence recover the predicted laws, while
our MHD numerical flows follow them in planes at lowest parallel wavenumbers. 

Financial support from PNST/INSU/CNRS are gratefully acknowledged. Computations were 
performed at IDRIS (CNRS) Grant No. 0597.

\end{document}